\documentclass[journal]{IEEEtran}

\usepackage{cite}
\usepackage{amsmath}
\usepackage{graphicx}
\usepackage{multirow}
\usepackage{color}
\usepackage{url}

\begin{document}

\title{Dataset artefacts in anti-spoofing systems: a case study on the ASVspoof 2017 benchmark} 

\author{Bhusan~Chettri,~\IEEEmembership{Student Member,~IEEE,}
        Emmanouil~Benetos,~\IEEEmembership{Senior Member,~IEEE}
        \\and~Bob~L.~T.~Sturm,~\IEEEmembership{Member,~IEEE} 
        \thanks{B. Chettri and E. Benetos are with the School of Electronic Engineering and Computer Science, Queen Mary University of London, UK. e-mail: \{b.chettri, emmanouil.benetos\}@qmul.ac.uk.}
        \thanks{B. L. T. Sturm is with the Division of Speech, Music and Hearing, KTH Royal Institute of Technology, Stockholm, Sweden. e-mail: bobs@kth.se.}
}

\markboth{IEEE/ACM Transactions on Audio, Speech, and Language Processing}%
{Shell \MakeLowercase{\textit{et al.}}: Bare Demo of IEEEtran.cls for TASLP}

\IEEEtitleabstractindextext{%
\begin{abstract}
The Automatic Speaker Verification Spoofing and Countermeasures Challenges motivate research in protecting speech biometric systems against a variety of different access attacks. The 2017 edition focused on replay spoofing attacks, and involved participants building and training systems on a provided dataset (ASVspoof 2017). More than 60 research papers have so far been published with this dataset, but none have sought to answer why countermeasures appear successful in detecting spoofing attacks. This article shows how artefacts inherent to the dataset may be contributing to the apparent success of published systems. We first inspect the ASVspoof 2017 dataset and summarize various artefacts present in the dataset. Second, we demonstrate how countermeasure models can exploit these artefacts to appear successful in this dataset. Third, for reliable and robust performance estimates on this dataset we propose discarding nonspeech segments and silence before and after the speech utterance during training and inference. We create speech start and endpoint annotations in the dataset and demonstrate how using them helps countermeasure models become less vulnerable from being manipulated using artefacts found in the dataset. Finally, we provide several new benchmark results for both frame-level and utterance-level models that can serve as new baselines on this dataset.

\end{abstract}

}

\maketitle

\IEEEdisplaynontitleabstractindextext
\IEEEpeerreviewmaketitle

\section{Introduction}\label{sec:introduction}

\IEEEPARstart{A}{utomatic} speaker verification (ASV) \cite{reynolds_SC1995} aims at verifying the claimed identity of a person using their voice characteristics. It is among the most convenient means of biometric authentication, but its robustness and security in the face of \emph{spoofing attacks} (or \emph{presentation attacks}) is of growing concern \cite{sahid_PAD_book}. Among various spoofing attack points identified in the ISO/IEC 30107-1 standard \cite{isopad}, the first two attack points: \emph{physical access} (PA) and \emph{logical access} (LA) attacks are of specific interest to researchers as these attacks enable an adversary to inject spoofed biometric data. PA attacks involve spoofed speech to pass through an ASV system's microphone. Impersonation \cite{lau_2004} and replay \cite{wu_APSIPA2014} are examples of such attacks. In contrast, LA attacks bypass the microphone by injecting computer generated speech or real stolen speech directly into an ASV system. Text-to-Speech \cite{Masuko99onthe} and Voice Conversion \cite{PellomH99} techniques are often used to produce artificial speech to perform LA attacks. High-stakes ASV applications \cite{lee2013speaker}, therefore, require an anti-spoofing system (countermeasures) to ensure fail-safe mechanisms. Most \textit{countermeasure} systems are defined as a binary classifier comprising a feature extractor and a backend that aims at discriminating \emph{bonafide} (human) speech from spoofing attacks.

In this paper, we study replay attacks using the ASVspoof\footnote{ASVspoof is an ASV community-driven effort promoting anti-spoofing research through open spoofing challenges releasing standard evaluation protocols, metrics and datasets. See \cite{wu_IS2015}, \cite{tomiSummaryPaper} and \cite{asvspoof2019overview} for an overview of the ASVspoof challenges held in 2015, 2017 and 2019 respectively.} 2017 dataset. The dataset has two different versions. Version 1.0 was used in the 2017 evaluations \cite{tomiSummaryPaper}. Post-evaluations, in \cite{bhusanICASSP2018} we demonstrated that initial zero-valued silence frames appearing in bonafide audio signals serve as a cue that Gaussian Mixture Model (GMM) based countermeasures exploit to form decisions. Subsequently, an updated version 2.0 dataset \cite{hectorAsvspoof2.0} was released by the ASVspoof organisers fixing these problems. Furthermore, our recent work \cite{bhusanSLT2018} on version 2.0 of the same dataset suggests that it contains some recording artefacts that impact performance estimates. In \cite{bhusanSLT2018} we analyse a state-of-the-art Convolutional Neural Network (CNN) based countermeasure model and find that it makes many decisions based heavily on the first few milliseconds. Further analysis of a few confidently classified test audio signals shows dual tone multi-frequency (DTMF) sounds or speech signals occurring in the first few milliseconds for spoofed audio, but nonspeech or silence in case of the bonafide audio signals. We hypothesize that the training and development subsets of the same dataset might contain such DTMF sounds and other confounders/artefacts that might influence model decisions. Understanding their statistics enables building trustworthy countermeasures using this dataset.

Spoofing countermeasures are mostly developed either in a frame-level \cite{hector_cqcc} or a fixed-duration utterance-level \cite{galina_IS2017} setting. As nonspeech samples may carry discriminative information such as differences in acoustic conditions, designing countermeasures using the entire audio signal is often a sensible design choice \cite{wu_IS2015,tomiSummaryPaper,asvspoof2019overview,galina_IS2017}. Since the release of the ASVspoof 2017 dataset, a number of frontend features \cite{hector_cqcc,patel_IS2017,buddhi_IS2018,tharshini_IS2018} and backend models \cite{fourthbestsystem_2017challenge,resnet_dataAugmentation,galina_IS2017,laiICASSP2019} have been proposed and studied. \emph{Constant Q cepstral coefficients} (CQCCs) \cite{hector_cqcc} that were initially proposed for synthetic and voice-converted spoofing detection have now been studied for replay detection and show encouraging results. \emph{Teager energy operator} (TEO) based spoof detection features were studied in \cite{patel_IS2017}. Speech demodulation features using the TEO and the Hilbert transform have been studied in \cite{madhu_IS2018}. Authors in \cite{buddhi_IS2018} proposed features for spoofing detection by exploiting the long-term temporal envelopes of the subband signal. Spectral centroid based frequency modulation features were proposed in \cite{tharshini_IS2018}. Some of the well-known deep models from computer vision including \emph{ResNet} \cite{he2015deep} and \emph{light CNN} \cite{wu2015light} were adapted for replay spoofing detection in \cite{fourthbestsystem_2017challenge,resnet_dataAugmentation} and \cite{galina_IS2017}, respectively demonstrating good performance on the ASVspoof 2017 dataset. Furthermore, \emph{attention-based models} for replay detection have also been studied in \cite{Tom2018} and \cite{laiICASSP2019}. Some of the new techniques that have emerged following the recent ASVspoof 2019 evaluations include use of \emph{SincNet} \cite{sincnet_asv_raw_audio} architecture which was originally developed for speaker recognition. Authors in \cite{but2019challenge} have used \emph{SincNet} for spoofing detection demonstrating good performance. Similarly, x-vectors \cite{x-vector_ASV} which were originally proposed for speaker recognition, have been studied for spoofing detection in \cite{jennifer2019challenge}.

However, most of these works focus on improving recognition performance without trying to understand what anti-spoofing systems are learning to make predictions. Systems showing better performance do not necessarily mean they are \textit{trustworthy} \cite{rusak2020increasing,jose_clever_hans}. To that end, following the guidelines from \cite{AIHLEG}, this paper considers technical robustness, fairness and accountability as key criteria for any machine learning (ML) model to be called \textit{trustworthy}. These criteria suggest that the results produced by a trustworthy model should be independent of variables/factors (in the training data) not relevant to the actual problem.

ML models make decisions from relationships discovered in training data \cite{bishop_ML}. As demonstrated in \cite{gtzan_dataset_faults,tommasi2015deeper,rosset_medical_datamining}, models can learn irrelevant cues, artefacts or confounders during training. Unless explicitly accounted for during training and inference such artefacts often contribute in achieving good results overestimating the actual performance on a testing set \cite{gtzan_dataset_faults,tommasi2015deeper,francisco_tismir,charalambous2016data,mendelson_medicine_ml_bias,Kaufman_data_leakage,rosset_medical_datamining,bhusan2019challenge,dan_automatic_acoustic_id}. In \cite{gtzan_dataset_faults}, a benchmark music informatics dataset is shown to have several faults that significantly bias performance metrics. In \cite{francisco_tismir}, the authors demonstrate how top performing music labeling systems were exploiting characteristics from the music signal that could not even be heard. In \cite{rosset_medical_datamining}, authors found the encoding of patient identity in a dataset correlates highly with cancer likelihood. The accuracy of a gait recognition system was found to drop when confounding factors were removed during training \cite{charalambous2016data}. Authors in  \cite{mendelson_medicine_ml_bias} describe how bias introduced as a result of dataset selection influenced the performance of an Alzheimer’s disease classification system. In computer vision applications, \cite{tommasi2015deeper} have analyzed the effect of biases in model performance across several datasets. In \cite{dan_automatic_acoustic_id}, the authors focus on building robust system for acoustic individual identification while reducing the effects of confounding factors in a dataset. Data leakage can often lead to overestimation of model performance, producing too-good-to-be-true results \cite{Kaufman_data_leakage}. One relevant work in this context in ASV anti-spoofing is that of \cite{Tom2018} who reported 0\% EER on both the development and evaluation sets of the ASVspoof 2017 v1.0 dataset --- the dataset on which the issue of silence providing cues \cite{bhusanICASSP2018} for class prediction was well acknowledged by the community \cite{hectorAsvspoof2.0}. 

The trustworthiness of models trained on data containing artefacts is therefore called into question; some can behave much like a ``horse'' \cite{jose_clever_hans, bobHorsePaper}, i.e. a model that uses irrelevant cues to make decisions \cite{sturm2016horse,francisco_tismir}. As highlighted in \cite{rosset_medical_datamining}, such confounders can occur as a result of data collection, compilation and partition. Such biases can have a severe impact on the trustworthiness of ML applications, which can be catastrophic for domains such as ASV anti-spoofing. Therefore, it is essential to perform an in-depth dataset analysis \cite{torralba_unbiased_look_dataset,Jonathan_ml_medicine}, detect presence of artefacts or confounders \cite{dan_automatic_acoustic_id}, ensuring models do not exploit irrelevant factors during training, and therefore yield reliable performance estimates.

\begin{figure}[t!]
	\centering  
	\includegraphics[width=\linewidth]{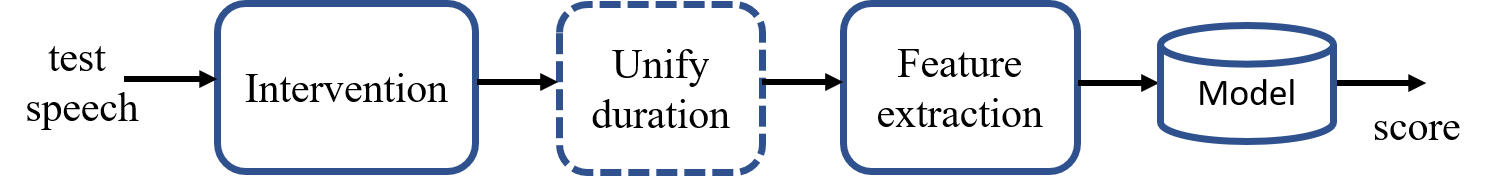}
	\caption{Block diagram: intervention pipeline towards understanding the influence of artefacts on the predictions of countermeasure models.}
	\label{intervention_pipeline}
\end{figure}

This paper extends our prior work \cite{bhusanSLT2018} from multiple perspectives and makes the following contributions:

\begin{itemize}
    
    \item We analyze the ASVspoof 2017 v2.0 dataset, discover different artefacts and summarize them in Section \ref{artefacts_in_dataset}. We investigate how these artefacts influence model decisions on five different countermeasures (Section~\ref{our_systems}) using interventions\footnote{Following \cite{francisco_tismir}, we define \textbf{\emph{Intervention}} as a process that updates audio signal either by adding or removing raw samples. Fig. \ref{intervention_pipeline} illustrates this idea.} (Section~\ref{exploring_impacts_setup1}). 
    
    \item We demonstrate how such artefacts can be used to attack countermeasures trained on this dataset, and emphasize why accountability of such artefacts while training countermeasures is important on this dataset (Section \ref{manipulating_models_decisions}).
    
    \item We propose a design framework (Fig. \ref{proposed_countermeasure_solution}) for trustworthy countermeasures on this dataset that incorporates speech endpoint detection during training and testing. Using this we propose a robust frame-based countermeasure and demonstrate its effectiveness against test signal manipulation (Section~\ref{overcoming_artefacts}). 
    
    \item We provide new benchmark results for frame-based and utterance-based countermeasures trained on constant Q-cepstral coefficient (CQCC) benchmark features, i-vectors and spectrograms. (Section~\ref{overcoming_artefacts}).
    
    \item We develop manual endpoint speech annotations, and make them (along with artefact file lists) publicly available in \cite{bhusan_chettri_2020_3601188}. Our source code is available online\footnote{https://github.com/BhusanChettri/TASLP-study-on-dataset-artefact}. We provide a supplementary document in \cite{bhusanTASLP_supplementary} for additional details of our work.

\end{itemize}

\section{Artefacts in the ASVspoof 2017 v2.0 dataset} \label{artefacts_in_dataset}

We perform a qualitative analysis\footnote{We use Audacity (https://www.audacityteam.org/) for visualizing and listening audio waveforms.} on all audio recordings in the training set ($1507$ bonafide and $1507$ spoof) and the development set ($760$ bonafide and $950$ spoof) of the ASVspoof $2017$ v2.0 dataset. As for the evaluation set, due to the large number of spoof recordings ($12008$), we perform this analysis only on the $1298$ bonafide recordings in this study. Furthermore, learning and optimizing model parameters requires use of the training and development sets, and thus we focus more towards understanding audio recordings in these two sets. For additional details on the dataset, please see \cite{hectorAsvspoof2.0}. Below we list our key observations that we consider unnatural or unexpected for a dataset. This could be due to faults made during data collection, compilation and post-processing. As we demonstrate in Section \ref{exploring_impacts_setup1}, some of them have a profound impact on model decisions raising concerns on the validity of results reported in the literature \cite{galina_IS2017,Tom2018,saranya_IS2018,gajan_IS2018}. \\

\noindent
\textbf{Pattern difference}. We define this as the \textit{presence} of \textit{nonspeech (noise, music or silence)} in the first $300$ milliseconds of bonafide recordings but missing in the spoof class recordings. About $60.45$\%, $73.55$\% and $69.1$\% of the bonafide audio files in the training, development and evaluation sets respectively have nonspeech. On the contrary, $68.74$\% and $41.05$\% of the spoof files in the training and development sets respectively have speech occurring within the first $300$ ms.  

\noindent
\textbf{Burst click sound (BCS).} We define this as an \textit{abrupt click sound} (low or loud) found in the start of audio recordings. About $36.36$\%, $23.55$\% and $41.06$\% of bonafide audio files in the training, development and evaluation sets were found to contain BCS in the start. On the contrary, $2.45$\% spoof files in the training set have BCS. No spoof class audio files in the development set have such BCS, and we do not have ground truth annotations for spoofed signals in the evaluation set.

\noindent	
\textbf{Dual-tone multi-frequency signaling (DTMF) sound.} About $45.58$\% ($687$ out of $1507$) of spoof audio files in the training set and $16.63$\% ($158$ out of $950$) in the development set were found to contain a DTMF sound (low or loud) within the first $200$-$250$ ms. The DTMF sound often overlaps with the actual spoken speech. We find 33.77\% ($232$ out of $687$) spoof files and 6.96\% ($11$ out of $158$) in the training and development sets contain such overlapping sounds. On the contrary, the bonafide class audio files do not have such DTMF sounds. 

\noindent
\textbf{Silence.} We find some bonafide audio recordings with more than $10$ ms zero valued silence in their start. There are $19.11$\% ($288$ out of $1507$) such bonafide files in the training set, $1.97\%$ ($15$ out of $760$) in the development set and $10.09$\% ($131$ of $1298$) in the evaluation set. Furthermore, in the training set we find $23.61$\% ($68$ out of $288$) files have more than $70$ ms silence and $12.85$\% ($37$ out of $288$) with more than $100$ ms silence in the start. In contrast no spoof class files are found to have such zero valued silence.

\noindent
\textbf{Corrupted audio files.} We find four audio recordings 
in the dataset that do not contain any speech at all.

\noindent
\textbf{Sentence S02 - ``Ok Google''.} It is one of the phrases used in the ASVspoof 2017 dataset with an average duration between $0.7$ - $0.8$ seconds. We find $165$, $136$ and $1282$ audio examples of S02 in the training, development and evaluation sets with more than $1.5$ seconds duration. This suggests that more than half of the contents of each recording contain noise or nonspeech.

\section[systems]{Spoofing countermeasures} \label{our_systems}
This section describes various countermeasures (features and classifiers) considered in this study to evaluate the influence of artefacts (Section \ref{artefacts_in_dataset}) on model decision. A description of evaluation metrics and initial model performance is also provided.

\subsection{Features and classifiers}
We study five different types of classifiers: GMMs trained on CQCCs \cite{hector_cqcc}; Cosine Distance and Support Vector Machines (SVMs) trained on CQCC-based i-vectors \cite{dehak_TASLP2011}; and two CNNs trained on time-frequency representations (spectrograms). The main motivations for choosing them are: (1) CQCCs coupled with GMMs (or deep neural networks) have been studied extensively on the ASVspoof 2017 dataset \cite{hectorAsvspoof2.0,replay_using_highFreqFeats,rohan_eCQCC,bhusanICASSP2018}; (2) CQCC-based i-vectors with Cosine backend classifiers have been used as a baseline system \cite{hectorAsvspoof2.0}; (3) spectrogram features with a CNN backend have shown the best performance during the ASVspoof 2017 challenge \cite{galina_IS2017}.

Furthermore, we aim to demonstrate that the artefacts outlined in Section \ref{artefacts_in_dataset} can affect any machine learning (ML) model and that the issues discussed do not revolve around a specific ML model.\\ 

\noindent
\textbf{GMM}: We train one GMM each for the bonafide and spoof classes using $512$ mixture components with random initialisation. We use $60$ dimensional CQCC features extracted using the setup from \cite{hectorAsvspoof2.0} to train the GMMs. During testing, for each test utterance $X$ (with $T$ feature vectors) a score is obtained using the log-likelihood ratio:
\begin{equation}
     \Lambda(X) = \frac{1}{T} \sum_{t=1}^{T} \log P(x_t|\theta_b) -  \frac{1}{T} \sum_{t=1}^{T} \log P(x_t|\theta_s)
\end{equation}
where $x_t$ is the $t^{th}$ frame, $P$ denotes the likelihood function, $\theta_b$ and $\theta_s$ represents the bonafide and spoof GMMs respectively. \\

\noindent
\textbf{Cosine}: We compute $100$-dimensional i-vectors using the same $60$ dimensional CQCC features used in GMMs for the entire dataset. We compute the mean i-vector corresponding to the bonafide and spoof classes in the training set and use them as the representative models. During testing, a similarity score is computed between a test i-vector and the model i-vector using the cosine distance metric:
\begin{equation}
\cos (\theta)  =  \frac{\mathbf{X} \cdot \mathbf{Y}}{||\mathbf{X}|| ||\mathbf{Y}||}\
\end{equation} where \textbf{X} represents the test i-vector and \textbf{Y} the model i-vector. The final score is then obtained by taking the difference between the bonafide and spoof model scores. We follow the same i-vector setup from \cite{hectorAsvspoof2.0}. \\

\noindent
\textbf{SVM}:
We train a binary SVM classifier with a linear kernel using mean-variance normalised i-vector features, with mean-variance values computed on the training set. Here, we use the same $100$ dimensional CQCC-based i-vectors used in the Cosine model. We use the Scikit-learn \cite{scikit-learn} library with default parameters for training and testing the SVM model. \\

\noindent
\textbf{CNN}:
We train two different CNNs: CNN$_1$ and CNN$_2$. Both CNNs operate on a fixed input representation. While CNN$_1$ works on $4$ seconds spectrogram input, CNN$_2$ operates on $3$ seconds duration. CNN$_1$ uses the architecture adapted from the best performing model \cite{galina_IS2017} of the ASVspoof 2017 challenge. CNN$_2$ is our proposed architecture with fewer model parameters comprising only 4 hidden layers in contrast to CNN$_1$ which is deep consisting of $9$ convolutional layers. The main motivation of using CNN$_2$ is twofold. First, we wanted to see how a simple CNN model (less deep) compares with the state-of-the-art adapted CNN of \cite{galina_IS2017}. Second, how different input feature configurations impact model performance. In CNN$_1$ the input spectrogram is computed using a $1728$ point FFT and a $108$ ms frame window with 10 ms hop size, while in CNN$_2$ we used a standard $512$ point FFT and $32$ ms frame window. We provide further details on CNN architecture, training and scoring in our supplementary document \cite{bhusanTASLP_supplementary}.

\subsection{Figures of merit}

We use the equal error rate (EER), the evaluation metric used in the ASVspoof 2017 challenge, to report the countermeasure performance. The EER defines an operating point where the false acceptance rate (FAR) and false rejection rate (FRR) of the system are equal. We further report FAR and FRR for the bonafide class to derive more insights in understanding the impact of the artefacts on this dataset:

\begin{equation}
\text{FAR} =  \frac{\text{FP}} {\text{FP + TN}}
\end{equation} 
\begin{equation}
\text{FRR} =  \frac{\text{FN}} {\text{TP + FN}}
\end{equation} 
where TP, TN, FP, and FN denote true positive, true negative, false positive and false negative counts respectively. Throughout the paper, we use the bonafide class as positive and the spoof class as negative in computing the above metrics.

\subsection{Initial model performance} \label{countermeasure_results_setup1}
We train all our countermeasure models using the training set and validate them on the development set. Table \ref{results_original_systems} summarizes the results. Countermeasures Cosine and GMM show consistent performance as reported in \cite{hectorAsvspoof2.0}. Although CNN$_2$ and CNN$_1$ show similar performance on the development set, CNN$_2$ performs poorly on the evaluation set. A possible reason could be due to the simple $4$ hidden layer architecture used by CNN$_2$ in comparison to the $10$ hidden layer representation in CNN$_1$. However, CNN$_2$ outperforms both Cosine, GMM and SVM on both the development and evaluation sets.

\begin{table}[th]	
	\caption{Initial model performance on the  development and evaluation sets. $\Theta$ = EER decision threshold.} 
	\centering
	\scalebox{0.96}{
		\begin{tabular} {cccccccc}	
			\hline
			Model & Set & $\Theta$ &TP & FN & FP & TN & EER\% \\			
			\hline
			\multirow{2}{*}{CNN$_1$} &Dev &$0.6663$ &$701$ &$59$ &$74$ &$876$ &$7.7$  \\
			&Eval &$0.7467$ &$1159$ &$139$ &$1286$ &$10722$ &$10.7$  \\
			\hline
			\multirow{2}{*}{CNN$_2$} &Dev &$0.6$ &$704$ &$56$ &$70$ &$880$ &$7.37$ \\
			&Eval &$0.842$ &$1124$ &$174$ &$1609$ &$10399$ &$13.4$ \\
			\hline
			\multirow{2}{*}{Cosine} &Dev &$0.125$ &$679$ &$81$ &$101$ &$849$ &$10.6$ \\
			&Eval &$0.181$ &$1105$ &$193$ &$1779$ &$10228$ &$14.8$ \\
			\hline
			\multirow{2}{*}{SVM} &Dev &$0.3972$ &$678$ &$82$ &$103$ &$847$ &$10.8$ \\
			&Eval &$0.506$ &$1094$ &$204$ &$1883$ &$10125$ &$15.6$ \\
			\hline
			\multirow{2}{*}{GMM} &Dev &$0.3334$ &$690$ &$70$ &$87$ &$863$ &$9.2$ \\
			&Eval &$0.7120$ &$1119$ &$179$ &$1656$ &$10352$ &$13.7$ \\
			\hline
		\end{tabular}}
		\label{results_original_systems}
\end{table}

In the next two sections \ref{exploring_impacts_setup1} and \ref{manipulating_models_decisions}, we show how the decisions of these models can be compromised using the artefacts described in Section \ref{artefacts_in_dataset}. Furthermore, it should be noted that we use the same $\Theta$ shown in Table~\ref{results_original_systems} (for all the models) to perform the intervention experiments described in sections \ref{exploring_impacts_setup1} and \ref{manipulating_models_decisions}. One motivation for this is to demonstrate a situation where a countermeasure trained on a biased dataset is being used to protect an ASV system leaving loopholes for being manipulated using dataset cues. Ensuring the training dataset to be free from such artefacts and biases is crucial towards designing a reliable spoofing detection system.

\section{Exploring the impact of dataset artefacts}\label{exploring_impacts_setup1}

In this section we thoroughly study the impact of artefacts (Section \ref{artefacts_in_dataset}) on countermeasure performance. More precisely we focus on understanding the influence of pattern difference, BCS and DTMF sounds on model prediction through intervention experiments illustrated in Fig.~\ref{intervention_pipeline}. We call this setup as \emph{inference-time intervention} because we use our pretrained models from Section \ref{our_systems}. Here the intervention module updates the test signal by exploiting information about the dataset artefacts which we pass as side information. Features are then computed on the updated test signal and scoring is performed using a pretrained model. The optional \emph{unify-duration} module truncates or replicates audio samples to create a fixed-duration input representation. This is applicable only for the CNNs.

\begin{figure*}[htb]
	\centering
	\begin{tabular}{@{}cc@{}}
		\includegraphics[width=0.48\textwidth]{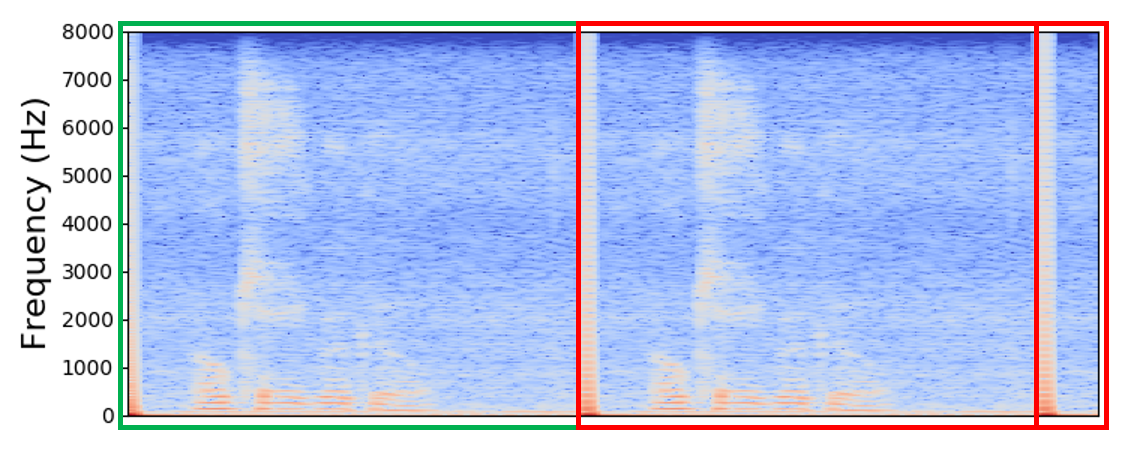} &
		\includegraphics[width=.48\textwidth]{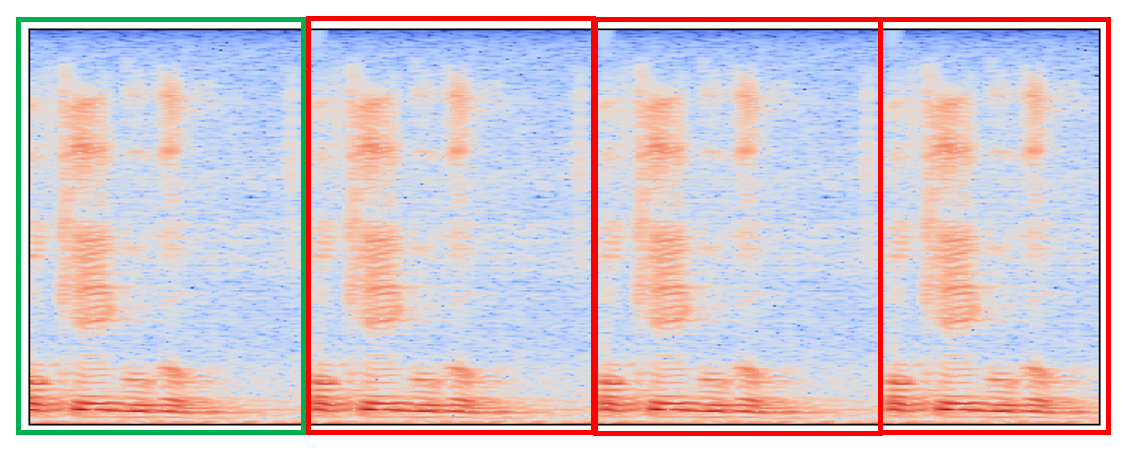} \\
		\includegraphics[width=0.48\textwidth]{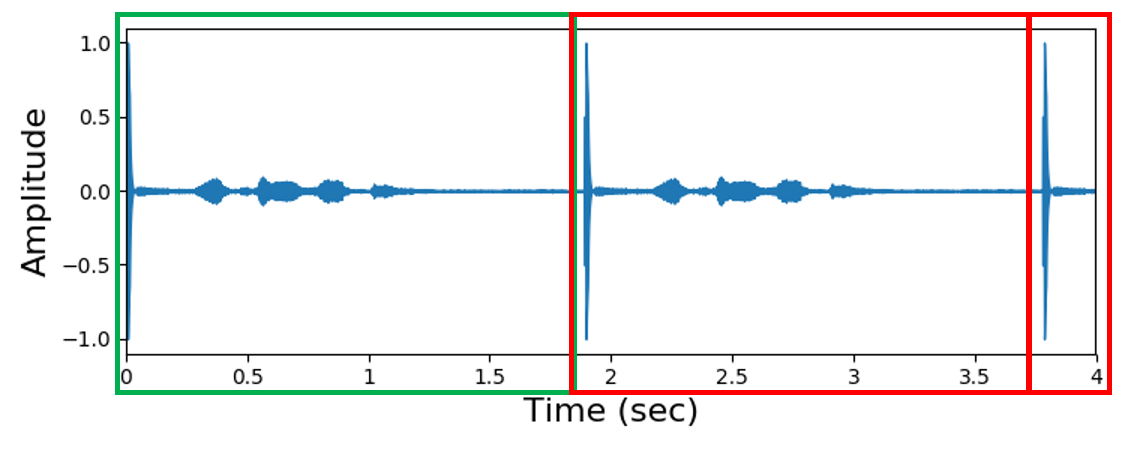} &
		\includegraphics[width=.48\textwidth]{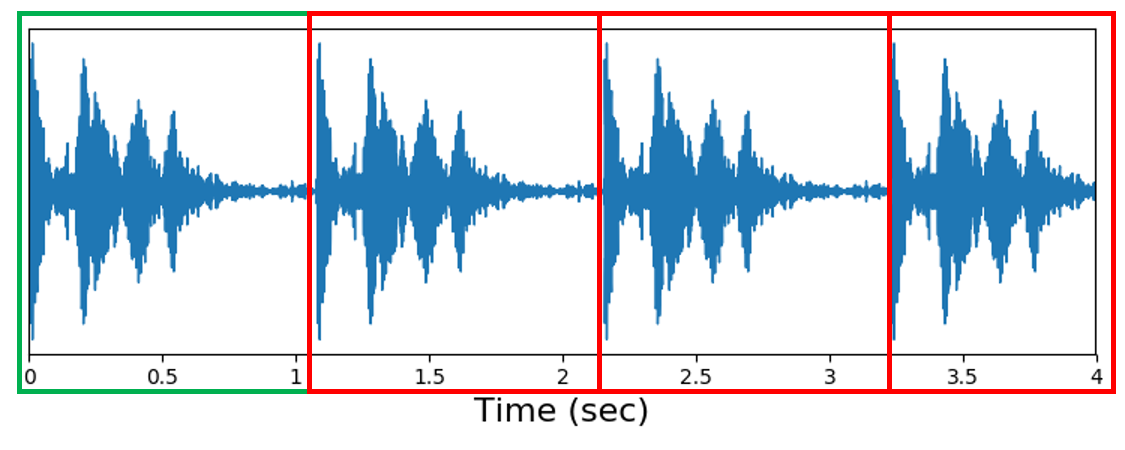}
		
	\end{tabular}
	\caption{Spectrogram (top) and raw audio (bottom) of ``Ok google''. Left represents a bonafide example and right its replayed version. The green rectangular box highlights the original audio and the red box shows signal replication to create a fixed duration (4 seconds) input representation. Two use cases are reflected: pattern difference and BCS. It shows how artefacts spread while creating an input representation with fixed duration.}
	\label{spectrogram_artefacts}
\end{figure*}

\begin{table}[th]
	\caption{BCS intervention results. TFI: test files intervened, which corresponds to TP cases (Table~\ref{results_original_systems}) identified to contain BCS artefact. Prop: proportion of files that changed class label.} 
	
	\centering
	\scalebox{1.0}{
		\begin{tabular} {ccccc}
			\hline			
			Model & Set &\# TFI &FN  &Prop (\%)  \\
			\hline
			\multirow{2}{*}{CNN$_1$} &Dev &$177$ &+$34$  &$19.21$  \\
			                         &Eval &$513$ &+$118$  &$23.0$   \\
			\hline
			\multirow{2}{*}{CNN$_2$} &Dev &$175$ &+$12$  &$6.85$  \\
			                         &Eval &$508$ &+$60$  &$11.81$   \\
			\hline
			\multirow{2}{*}{Cosine} &Dev &$159$ &+$8$ &$5.03$ \\
			                        &Eval &$486$ &+$32$ &$6.58$ \\
			\hline
			\multirow{2}{*}{SVM} &Dev &$159$ &+$6$ &$3.77$ \\
			                     &Eval &$491$ &+$32$ &$6.51$ \\
			\hline
			\multirow{2}{*}{GMM} &Dev &$ 173$ &+$13$ &$7.51 $  \\
			                     &Eval &$508$ &+$56$ &$11.02$  \\
			\hline
		\end{tabular}}
		\label{bcs_genuine_results_summary_new}
\end{table}

\subsection{Impact of ``BCS'' on model prediction} \label{bcs_experiment}

The training set contains a large proportion ($36.36$\%) of bonafide examples with BCS artefacts in comparison to only $2.45$\% of spoof examples. BCS, if present in an audio recording, usually occurs within a 100 ms time window and is found either at the start or at the end. Although few (10.81\%) bonafide class audio files in the training set have BCS at the end, our preliminary interventions showed they have no impact on model predictions. However, we find a substantial influence for BCS found at the start of the audio recordings. We do not perform this intervention on the spoof class as the BCS serves as a cue for the bonafide class. Therefore we hypothesize and demonstrate that BCS serves as one kind of bonafide signature on this dataset. And if this signature is not removed, machine learning countermeasures can easily exploit it. To this end, we take all the TPs for the bonafide class that contain a BCS at the start and run this intervention on them. 

Here our intervention module (Fig.~\ref{intervention_pipeline}) takes a BCS annotation file containing a list of files having a BCS as side information. It then discards the first $100$ ms audio samples from the test utterance before extracting features and obtaining a classification score. Table \ref{bcs_genuine_results_summary_new} summarizes the results. As expected, dropping BCS samples causes models to misclassify bonafide test signals raising false negatives. Interestingly, we find CNN$_1$ to be more sensitive in contrast to CNN$_2$ and other models. A possible explanation is since CNN$_1$ uses a $4$ second representation in contrast to the $3$ seconds one for CNN$_2$. The above mentioned representation of CNN$_1$ contains more replicated copies of shorter audio segments, which propagates artefacts (see Fig.~\ref{spectrogram_artefacts}). 

\subsection{Impact of ``DTMF'' on model prediction}
		
During replay data collection a number of bonafide utterances were first concatenated using a DTMF sound to mark the utterance boundary and replay attacks were simulated on them. The individual replayed utterance was then retrieved based on this marker \cite{kinnunen2017reddots}. As outlined in Section~\ref{artefacts_in_dataset}, some spoof audio files in the training and development set have DTMF sounds (low or loud) which are not present in the bonafide files. DTMF, if present in an audio recording, usually occurs within a $250$ ms time window at the start.

Do DTMF sounds provide cues for the spoof class? Do these dataset artefacts bias our ML models? We perform interventions to understand this. As highlighted earlier in Section \ref{artefacts_in_dataset}, the ground truth of DTMF artefacts for the spoof class in the evaluation set is unavailable, and hence the present study does not include this intervention on them. To this end, we take all the TNs for the spoof class in the development set that contain a DTMF artefact and run this intervention. We pass the file identifier containing a DTMF as side information to the intervention module (Fig.~\ref{intervention_pipeline}) which removes the first $250$ ms audio samples from them before extracting features and obtaining a classification score.

\begin{table}[th]
	\caption{DTMF intervention results for the development set spoof files identified to contain a DTMF. Prop has the same meaning as in Table \ref{bcs_genuine_results_summary_new}. TFI: test files intervened.}

	\centering
	\scalebox{1.0}{
		\begin{tabular} {cccc}	
			\hline		
			Model & \# TFI &FP & Prop (\%) \\
			\hline
			\multirow{1}{*}{CNN$_1$} &$136$ &+$3$  &$2.2$  \\		
			\multirow{1}{*}{CNN$_2$} &$144$ &+$3$ &$2.08 $ \\
			\multirow{1}{*}{Cosine} &$145$ &$0$ &$0.0$ \\
			\multirow{1}{*}{SVM} &$145$ &$0$ &$0.0$ \\
			\multirow{1}{*}{GMM} &$141$ &+$1$ &$0.71$ \\
			\hline
		\end{tabular}}
		\label{dtmf_correctly_classified_spoof_interventions_summary}
\end{table}			
	
Table~\ref{dtmf_correctly_classified_spoof_interventions_summary} summarizes the intervention results. We see a negligible proportion of intervened files affected from this intervention, which suggests that DTMF sounds do not provide a substantial bias on model decisions. Another interpretation is the fact that the spoof signals acquire other channel characteristics during the replay simulation. Therefore, their impact may be negligible when audio signals are replayed in noisy acoustic conditions.  

\subsection{Impact of ``pattern difference'' on model prediction} \label{pattern_difference_intervention}

The previous two experiments involved removing BCS and DTMF artefacts from the test files identified to contain such artefacts. In this experiment we remove audio samples before and after the actual speech utterance during testing ensuring that both bonafide and spoofed audio recordings now have similar audio patterns. This also means that BCS or DTMF (if present) gets removed in this intervention experiment. Thus, BCS and DTMF experiments can be thought of as a special case of pattern difference interventions but performed on a small set of test files identified to contain such artefacts.

Here, the intervention module uses speech endpoints as side information and discards audio samples before and after the actual speech utterance. We use our manual speech endpoint annotations that we prepared during dataset inspection. Refer to \cite{bhusanTASLP_supplementary} for additional details on the annotations. Following our prior findings in \cite{bhusanSLT2018} we hypothesize that the pattern difference favours the bonafide class. To confirm this, we take all the TPs for the bonafide class and all the FPs for the spoof class (from Table~\ref{results_original_systems}) and run this intervention on them. As we did not perform qualitative analysis on spoof files in the evaluation set (Section \ref{artefacts_in_dataset}), the speech endpoint annotations for them are not available. Therefore, we do not run this intervention on evaluation set spoof files in the present study. 

\begin{table}[th]
	\caption{Pattern difference intervention results. To be compared with Table~\ref{results_original_systems}. `+', `-' denotes an absolute increase/decrease. Ground truth annotations for spoof files in the evaluation set are unavailable (indicated by $-$).} 
\centering
\scalebox{1.0}{
	\begin{tabular} {cc|cc|cc}	
		\hline
		\multicolumn{2}{c|}{Intervention on} & \multicolumn{2}{c|}{Bonafide class}  & \multicolumn{2}{c}{Spoof class} \\
		\hline
		Model & Set & FN & FRR \% & FP & FAR \% \\
		\hline
		\multirow{2}{*}{CNN$_1$} &Dev &+$334$  &+$43.95$ &-$49$  &-$5.16$  \\
		&Eval &+$519$  &+$39.98$ & $-$ & $-$  \\
		\hline
		\multirow{2}{*}{CNN$_2$} &Dev &+$73$  &+$9.61$ &-$35$ &-$3.68$ \\
		&Eval &+$289$  &+$22.27$ & $-$ & $-$ \\
		\hline
		\multirow{2}{*}{Cosine} &Dev &+$155$ &+$20.39$ &-$53$ &-$5.58$ \\
		&Eval &+$352$ &+$27.12$ &$-$ &$-$\\
		\hline
		\multirow{2}{*}{SVM} &Dev &+$174$ &+$22.89$ &-$52$ &-$5.47$ \\
		&Eval &+$349$ &+$26.89$ &$-$ &$-$ \\
		\hline
		\multirow{2}{*}{GMM} &Dev &+$170$ &+$22.37$ &-$41$ &-$4.32$ \\
		&Eval &+$429$ &+$33.13$ &$-$ &$-$ \\
		\hline
	\end{tabular}}
	\label{pattern_difference_genuine_spoof}
\end{table}

Table \ref{pattern_difference_genuine_spoof} summarizes the results of this intervention. As expected, a large number of bonafide test examples on both the development and evaluation sets are now misclassified by all our ML models as shown from the increased FN and FRR\% metrics. On the spoof class (development set) we find a drop in the FP and FAR\% metrics for all ML models. These results confirm our hypothesis about this pattern difference on this dataset. It indeed favours the bonafide class. This makes sense since a large proportion of bonafide audio files in the training set have silence/nonspeech in the first $300$ ms while the spoof class contains speech. 

\section{Attacking countermeasure models}
\label{manipulating_models_decisions}

In the previous section we demonstrated that BCS artefacts favour the bonafide class as evident from the increase in FRR\% when we removed them from the bonafide test files. We now show how easy it is to attack countermeasure models by using dataset artefacts as a class cue. More precisely, we perform a similar intervention as in Section \ref{exploring_impacts_setup1} with one difference. The intervention module now adds bonafide class cues to the test files. We study fooling countermeasures using three different bonafide class signatures: BCS, synthetic sound (mimicking BCS artefact), and zero-valued silence. We perform these interventions on all the \textit{misclassified bonafide files} (FNs) and \textit{correctly detected spoof files} (TNs) from Table \ref{results_original_systems}.

\subsection{Using BCS as a cue for the bonafide class}
Here the intervention module (Fig.~\ref{intervention_pipeline}) takes as side-information a ``BCS'' signature and appends it to the start of test audio recordings before passing on to the other modules for feature extraction and scoring. As a BCS signature we use the first $100$ ms samples of the most confidently classified bonafide audio ``T\_1001039.wav'' containing a BCS artefact in the training set. It should be noted that we did a similar line of study in \cite{bhusanSLT2018} and \cite{bhusanICASSP2018} but this study is different in terms of the signature we used for interventions. Furthermore our current study can be viewed as an extension of our prior work \cite{bhusanSLT2018}. In \cite{bhusanICASSP2018} we used $60$ ms zero-valued silence as a signature to fool GMM-based countermeasures on version 1.0 of this dataset. In \cite{bhusanSLT2018}, we find that CNNs give strong emphasis on the first $400$ ms for class discrimination. And, we used the initial $400$ ms samples as a signature to fool the prediction of the CNN countermeasure.

\begin{table}[th]
\caption{Manipulating model decisions using BCS. To be compared with Table~\ref{results_original_systems}. `+', `-' denotes an absolute increase/decrease.} 
	\centering
	\scalebox{1.0}{
	\begin{tabular} {cc|cc|cc}	
	\hline

    \multicolumn{2}{c|}{\multirow{2}{*}{Intervention on}} & \multicolumn{2}{c|}{Misclassified}  & \multicolumn{2}{c}{Correctly detected} \\
	& & \multicolumn{2}{c|}{bonafide files}  & \multicolumn{2}{c}{spoof files} \\
	\hline
	Model & Set & FN & FRR \% & FP & FAR \% \\
	\hline
	\multirow{2}{*}{CNN$_1$} &Dev  &-$46$  &-$6.05$ &+$446$ &+$46.95$  \\
	&Eval &-$129$  &-$9.94$ &+$4909$ &+$40.88$  \\
	\hline
	\multirow{2}{*}{CNN$_2$} &Dev  &-$56$  &-$7.37$ &+$479$ &+$50.42$ \\
	&Eval &-$153$ &-$11.79$ &+$4937$  &+$41.11$  \\
	\hline
	\multirow{2}{*}{Cosine} &Dev &-$37$ &-$4.87$ &+$130$ &+$13.68$ \\
	&Eval &-$54$ &-$4.16$ &+$1857$ &+$15.46$ \\
	\hline
	\multirow{2}{*}{SVM} &Dev &-$33$ &-$4.34$ &+$106$ &+$11.16$ \\
	&Eval &-$53$ &-$4.08$ &+$1952$ &+$16.26$ \\
	\hline
	\multirow{2}{*}{GMM} &Dev &-$51$ &-$6.71$ &+$325$ &+$34.21$ \\
	&Eval &-$142$ &-$10.94$ &+$5732$ &+$47.73$ \\
	\hline
	\end{tabular}}
	\label{bcs_misclassified_genuine_correctly_classified_spoof_results}
\end{table}
				
Table~\ref{bcs_misclassified_genuine_correctly_classified_spoof_results} shows the intervention results in terms of absolute increase/decrease in the error metrics. The consistency in the drop of FN and FRR\% and the increase in FP and FAR\% across all ML models confirm our hypothesis about BCS. It indeed serves as a strong cue that a model attends to form bonafide class decisions. The GMM and CNNs in particular show a high impact of this intervention on the evaluation set. For example $124$ misclassified bonafide files are now correctly classified by CNN$_1$ (out of $139$) and $142$ by the GMM (out of $179$). Furthermore FAR raises by more than $40$\% for the GMM and CNNs demonstrating that a large amount of correctly detected spoof files are now able to bypass them. Even though i-vectors are computed on CQCC features, the impact on Cosine and SVM models that operate on i-vectors is much smaller than GMMs. A possible reason for this is that i-vectors involve feature aggregation across all time frames during super-vector computation which may have reduced the influence of BCS on the final i-vector feature.

For further insights we provide visualizations of the original scores and scores obtained after the BCS intervention in our supplementary document \cite{bhusanTASLP_supplementary}.

\begin{table}[th]
\caption{Manipulating model decisions using white noise. To be compared with Table~\ref{results_original_systems}.} 
\centering
\scalebox{1.0}{
\begin{tabular} {ccc|cc|cc}	
\hline
\multicolumn{3}{c|}{\multirow{2}{*}{Intervention on}}& \multicolumn{2}{c|}{Misclassified}  & \multicolumn{2}{c}{Correctly detected} \\
&& & \multicolumn{2}{c|}{bonafide trials}  & \multicolumn{2}{c}{spoof trials } \\
\hline		
Model &SNR &Set &FN &FRR &FP &FAR\\ 
\hline
\multirow{4}{*}{CNN$_1$} & \multirow{2}{*}{0} &Dev &-$26$ &-$3.42$ &+$394$ &+$41.47$  \\
	& &Eval  &-$114$ &-$8.78$ &+$4583$ &+$38.17$  \\
	\cline{2-7}
	& \multirow{2}{*}{6} &Dev  &-$38$ &-$5.0$ &+$176$ &+$18.53$  \\
	&  &Eval   &-$94$ &-$7.24$ &+$1138$ &+$9.48$  \\
	\hline
		
	\multirow{4}{*}{CNN$_2$} & \multirow{2}{*}{0} &Dev  &-$43$ &-$5.66$ &+$662$ &+$69.68$  \\
	& &Eval  &-$136$ &-$10.48$ &+$6748$ &+$56.2$  \\
	\cline{2-7}
	& \multirow{2}{*}{6} &Dev  &-$40$ &-$5.26$ &+$164$ &+$17.26$ \\
	&  &Eval  &-$73$ &-$5.62$ &+$2113$ &+$17.6$ \\
	\hline
	\multirow{4}{*}{Cosine} & \multirow{2}{*}{0} &Dev  &-$63$ &-$8.29$ &+$161$ &+$16.95$  \\
	& &Eval  &-$78$ &-$6.01$ &+$1425$ &+$11.87$  \\
	\cline{2-7}
	& \multirow{2}{*}{6} &Dev  &-$13$ &-$1.71$ &+$14$ &+$1.47$ \\
	&  &Eval  &-$13$ &-$1.0$ &+$240$ &+$2.01$ \\
	\hline
	\multirow{4}{*}{SVM} & \multirow{2}{*}{0} &Dev  &-$61$ &-$8.03$ &+$139$ &+$14.63$  \\
	& &Eval  &-$74$ &-$5.70$ &+$1333$ &+$11.10$  \\
	\cline{2-7}
	& \multirow{2}{*}{6} &Dev  &-$11$ &-$1.45$ &+$20$ &+$2.11$  \\
	&  &Eval  &-$19$ &-$1.46$ &+$267$ &+$2.22$  \\
	\hline
	\multirow{4}{*}{GMM} & \multirow{2}{*}{0} &Dev  &-$17$ &-$2.24$ &+$46$ &+$4.84$  \\
	& &Eval  &-$20$ &-$1.54$ &+$660$ &+$5.50$  \\
	\cline{2-7}
	& \multirow{2}{*}{6} &Dev  &-$27$ &-$3.55$ &+$49$ &+$5.16$  \\
	&  &Eval  &-$23$ &-$1.77$ &+$368$ &+$3.06$  \\
	\hline
							
\end{tabular}}
\label{results_controlled_intervention_random_location_setup1}
\end{table}

\subsection{Using synthetic cues for the bonafide class}
What if an attacker does not have access to the BCS signature? Can they still fool countermeasure models trained on this dataset using a synthesized burst sound? To demonstrate this, we now repeat the same BCS interventions but use $100$ ms white noise as a signature to fool ML models. We experiment with synthetic noise at different signal to noise ratios (SNR) and demonstrate that white noise with enough power can fool ML decisions serving as a cue for the bonafide class. To ensure that the power of the original and manipulated speech signal is equivalent after adding noise, we first normalise the noise samples. Let $X_i$ be a test audio signal and $n_i$ represent the noise samples drawn from a standard normal distribution. Therefore the signal to noise ratio (SNR) can be written as:
\begin{equation}
\text{SNR} =  \text{log}_{10} \left[ \frac{\text{Var}(X_i)} {\text{Var}(\alpha \times n_i)} \right]\label{eq:snr}
\end{equation} 
where $\alpha$ is the scalar we want to compute for a given $X_i$ and an SNR and Var(.) represents variance. Thus:
\begin{figure}[t!]
	\centering  
	\includegraphics[width=\linewidth]{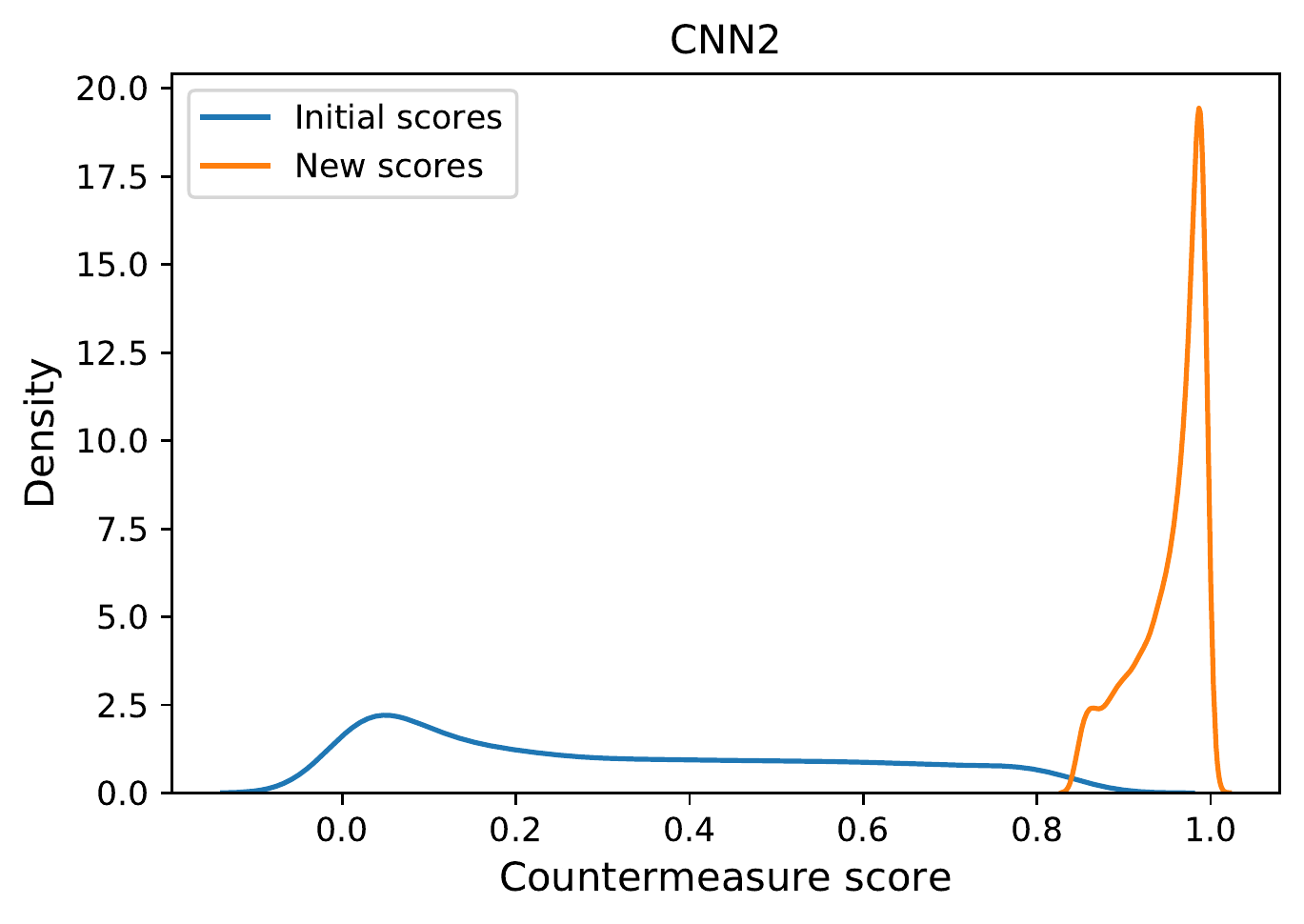}
	\caption[Score distributions of true negatives that gets misclassified after adding $100$ ms silence at random time locations.]{Score distributions of spoof files in the evaluation set that were originally detected correctly by CNN$_2$ and are now misclassified after adding $100$ ms silence at random time locations.} 
	\label{intervene_cnn2_spoof_silence_random_location}
\end{figure}

\begin{equation}
\alpha = \sqrt{\text{Var}(X_i) \times 10^{-\text{SNR}} } .\label{eq:alpha}
\end{equation} 

Using $\alpha$ in Eq.~\ref{eq:alpha}, the intervention module (Fig.~\ref{intervention_pipeline}) normalises the random noise before appending it to a test signal. The updated signal is then propagated for feature extraction and scoring. We investigate the impact of synthetic noise at different SNR levels and at the start and random time locations. In general, we find similar trends in the results for the evaluation set with interventions at different time locations. The impact reduces as we increase the SNR from 0 to 6, and CNNs are more sensitive in contrast to other countermeasure models. Furthermore, for CNNs we find the impact to be higher for interventions at the starting time location in contrast to adding synthetic noise at random time locations. For example, on the evaluation set, with SNR $0$, CNN$_1$ reports an FAR of 48.94\% when we added synthetic noise at the start and an FAR of 38.17\% for random time locations. An explanation to this accounts for the signal replication procedure used in CNN$_1$ to match the utterance duration to 4 seconds. As illustrated in Fig. \ref{spectrogram_artefacts}, adding artefacts (synthetic noise in this case) at the beginning would have more replication (for cases when original utterance duration is shorter than 4 seconds) of the artefact in contrast to adding it at random time locations, which explains the reason for the higher impact at starting time locations.

Therefore, in this paper we only include results for interventions at random time locations and provide results for the start time location in our supplementary document \cite{bhusanTASLP_supplementary}. Table~\ref{results_controlled_intervention_random_location_setup1} shows the intervention results. In general, on both the development and evaluation sets we find the impact to become less effective as we increase the SNR. A possible interpretation could be that a $100$ ms noise at 6 dB has smaller power/amplitude and does not exhibit a strong burst-sound property in contrast to a 0 dB noise. While this interpretation holds true for most of our ML models, we observe a slightly different behaviour for the GMM. An interpretation to this is that GMMs trained on version 2.0 of the dataset are still sensitive to silence (see \cite{bhusanTASLP_supplementary} for details) and $100$ ms noise samples drawn at $6$ dB SNR have low energy in contrast to noise samples at $0$ dB SNR. Although we see a similar trend (as in Table~\ref{bcs_misclassified_genuine_correctly_classified_spoof_results}) in FRR\% and FAR\%, we find much smaller impact for the GMM using noise in comparison to BCS (compare Table~\ref{bcs_misclassified_genuine_correctly_classified_spoof_results} and \ref{results_controlled_intervention_random_location_setup1}). A possible reason could be that we normalise the noise with respect to the original signal power (Eq.~\ref{eq:alpha}) which is not performed with BCS. We simply copy the BCS (raw samples containing BCS) and append it to the test signal during the intervention (Fig.~\ref{intervention_pipeline}). 

\subsection{Using silence as a cue for the bonafide class} \label{silence_cue}
As highlighted in Section \ref{artefacts_in_dataset}, some bonafide audio files in this dataset still contain zero-valued silence which is missing from spoofed files. Therefore, we run the intervention (Fig.~\ref{intervention_pipeline}) using silence as a bonafide class cue to fool our five countermeasures trained on this dataset. As we did a similar experiment on version $1.0$ of this dataset in \cite{bhusanICASSP2018}, we provide all the experimental details and results in our supplementary document \cite{bhusanTASLP_supplementary}. However, we provide an insight on the impact of this attack through score visualization for one of our countermeasure in Fig. \ref{intervene_cnn2_spoof_silence_random_location}. We observe that all the true negative examples are now misclassified as bonafide class confirming the hypothesis that silence indeed serve as potential cue on this dataset. Our supplementary document provides detailed explanation.

\begin{figure}[t!]
	\centering  
	\includegraphics[width=\linewidth]{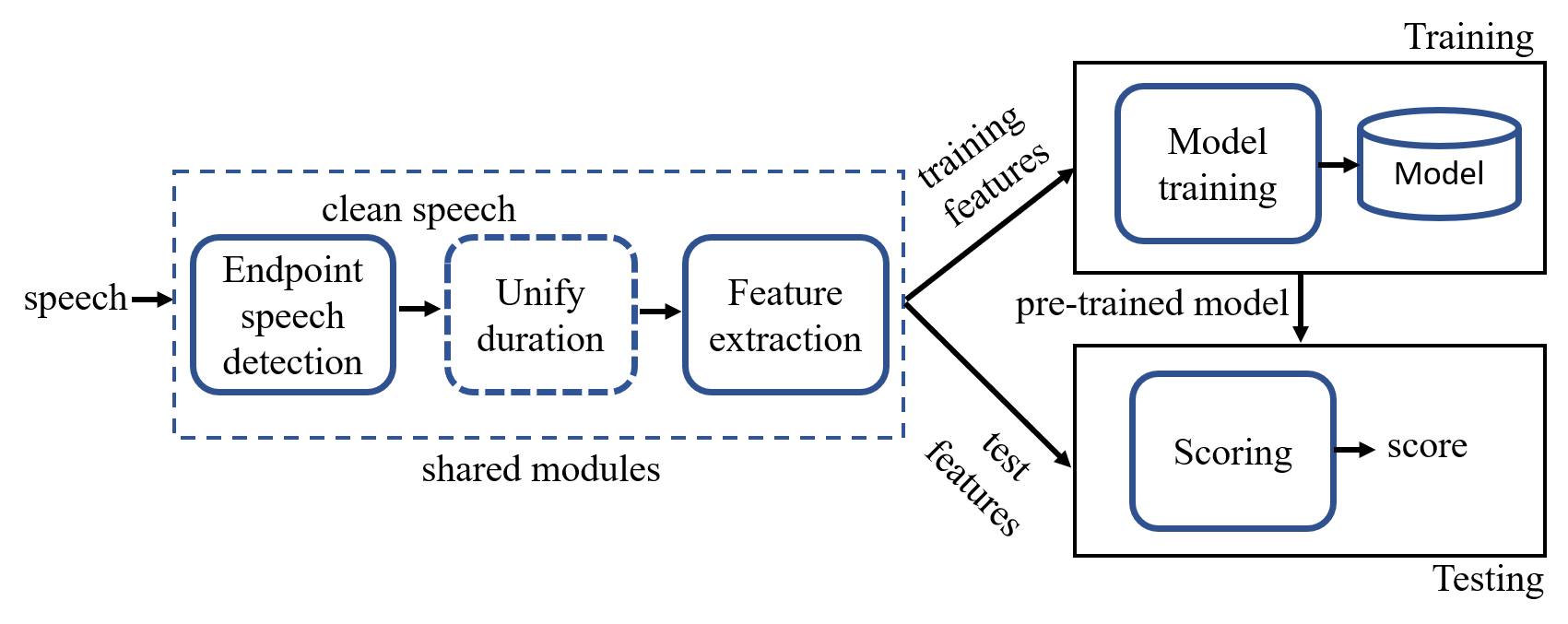}
	\caption{Proposed countermeasure design for reliable performance estimates.}
	\label{proposed_countermeasure_solution}
\end{figure}

\section{Overcoming the impact of dataset artefacts} \label{overcoming_artefacts}
We now describe our proposed methodology to address the issues highlighted in our study so far. To this end, we propose the use of speech endpoint detection during training and inference to build reliable and trustworthy countermeasures on this dataset. Fig.~\ref{proposed_countermeasure_solution} illustrates our proposed idea. The first three blocks are shared during training and testing. The endpoint detection module removes raw samples before and after the actual speech utterance\footnote{Our proposed speech endpoint detection approach (manual and automatic) do not remove nonspeech/silence within the utterance.}. This ensures that both bonafide and spoof utterances now have similar audio patterns and are free from recording artefacts we highlighted in Section \ref{artefacts_in_dataset}. The cleaned speech signal is then passed on to the subsequent modules for feature extraction, model training and inference. As before (Section \ref{exploring_impacts_setup1}) the role of the `unify duration' module remains the same. Furthermore, we now discard the use of corrupted audio files (Section~\ref{artefacts_in_dataset}) during training and testing.

In the following, we first describe the approach used for speech endpoint detection. We then explain our proposed frame-level deep countermeasure model for robust performance. Then we evaluate the performance of our new models trained with endpoint detection providing new benchmark results. Finally, we demonstrate that our newly trained models are more robust against being manipulated using cues (BCS artefacts) found in this dataset in contrast to the initial models trained without speech endpoint detection.

\subsection{Speech endpoint detection}

\begin{figure}[t!]
	\centering  
	\includegraphics[width=7cm]{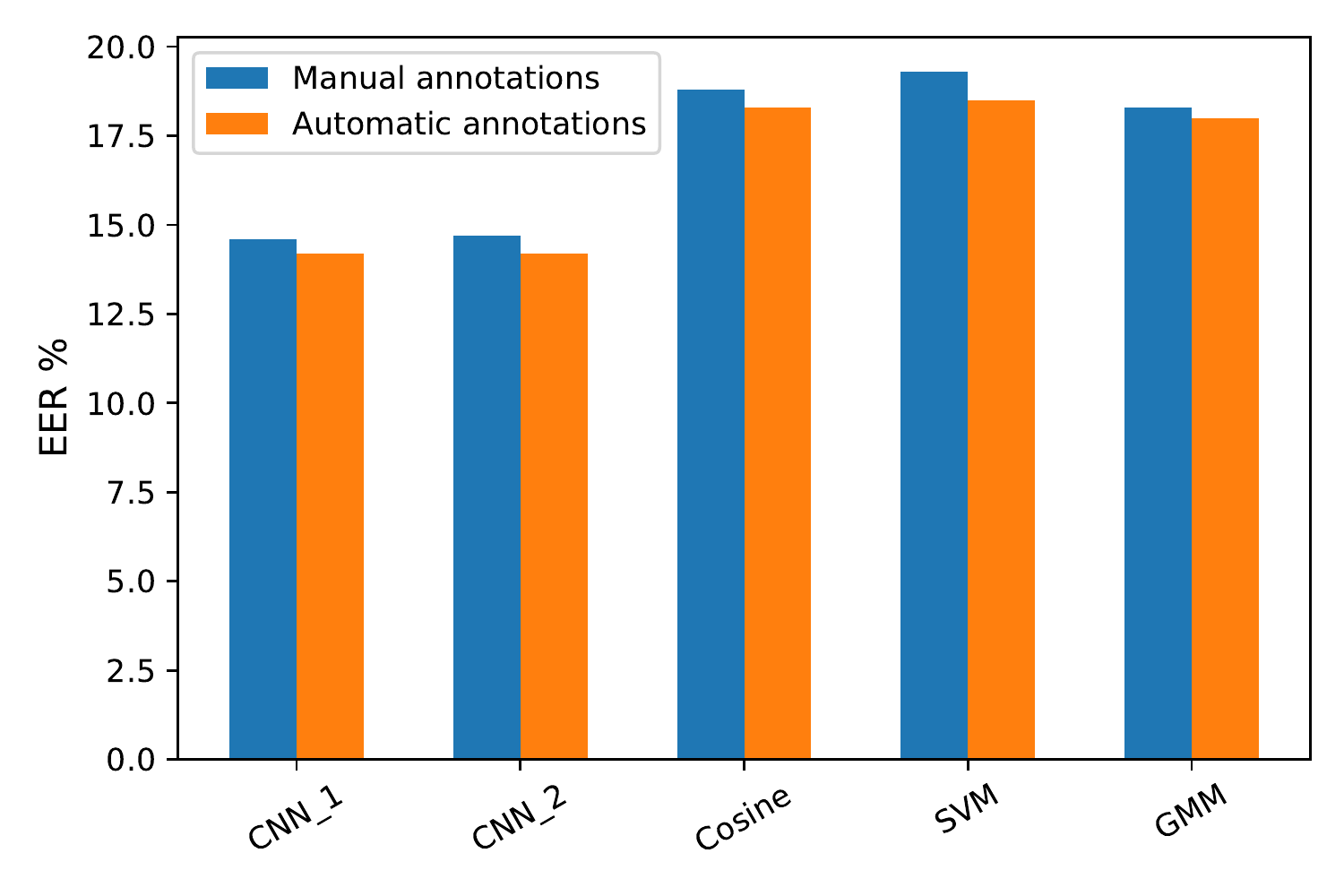}
	\caption{Performance (EER\%) on the evaluation set using models trained with manual and automatic speech endpoint annotations.}
	\label{barchart_manual_automatic_annotations_results}
\end{figure}

We use two approaches for speech endpoint detection: manual and automatic. The manual approach uses speech endpoint annotations that we collected during the dataset inspection. Automatic speech endpoint detection is based on rVAD \cite{TAN20201}, a robust voice activity detection algorithm. Please refer to \cite{bhusanTASLP_supplementary} for details on the two approaches. We use manual speech endpoint annotations for training and validating models and, use automatic endpoint detection during testing (as we do not have manual annotations). Automatic methods may yield some incorrect endpoint annotations. Therefore, it is important to first compare the accuracy of automatic and manual speech endpoint annotations. For this, we train (and validate) all our countermeasure models using both manual and automatic annotations and, use automatic annotations during testing. We present the results of this experiment in Fig. \ref{barchart_manual_automatic_annotations_results}. As can be seen, for all our countermeasure models we observe a comparable EER\% (on the evaluation set) between models trained using automatic and manual endpoint annotation. These results suggests that our proposed hypothesis of using manual annotations during training (and validation) but using automatic annotations during testing would not affect much on model prediction.

\subsection{Frame-level deep countermeasure model}
In this study we have found that neural network based countermeasures (CNNs) not only show superior performance over other countermeasures, but are equally more sensitive to artefacts in this dataset. One reason for this accounts to the fixed-duration input representation used by them. Duplicating audio contents to match the desired duration also involves spreading artefacts in the audio signal (see Fig. \ref{spectrogram_artefacts}). As a result they become more sensitive to artefacts (see Tables \ref{pattern_difference_genuine_spoof}, \ref{bcs_misclassified_genuine_correctly_classified_spoof_results}), and hence they are less trustworthy. Motivated from this, we propose a frame-level deep countermeasure model (DNN) that is trained on the original audio contents (i.e non-replicated or truncated), which is more reliable than CNNs. The use of context-frames --- augmenting past and future time frames to the current time frame --- is often adopted in training frame-level DNNs \cite{cai_IS2017}. However, for direct comparison with GMMs we do not use any context-frames in this work. This DNN treats the input as a bag-of-frames much like the way GMMs are trained. Its architecture comprises a series of fully connected layers and operates on a single input frame to predict whether the frame corresponds to a bonafide or spoofed class. We provide more details related to architecture, training and scoring of this DNN in our supplementary document \cite{bhusanTASLP_supplementary}.

For comparison and completeness of the study, we first train and evaluate this DNN on the original dataset (without endpoint detection). Our DNN reports an EER of $28.95$\% on the evaluation set which is worse than CNN$_1$ ($10.7$\%) and CNN$_2$ ($13.4$\%) trained using fixed-input representations. Although CNNs trained with context information yield better detection performance over the frame-based DNN, we later demonstrate (Section \ref{new_model_robustness_test}) that DNNs are more robust and trustworthy than CNNs.

\begin{table}[th]
	\caption{New benchmark results (EER \%) using models trained with speech endpoint detection. 1: no endpoint detection. 2: using automatic endpoint detection.}
	\centering
	\scalebox{1.0}{
		\begin{tabular} {cccc}
			\hline
			Model & Set & Test condition 1 & Test condition 2 \\
			\hline
			\multirow{2}{*}{CNN$_1$} &Dev  &$7.76$ &$9.0$  \\
			&Eval &$17.2$ &$14.58$ \\
			\hline
			\multirow{2}{*}{CNN$_2$} &Dev  &$8.6$ &$9.49$ \\
			&Eval  &$15.16$ &$14.77$\\
			\hline
			\multirow{2}{*}{Cosine} &Dev  &$14.76$ &$15.91$ \\
			&Eval &$20.49$ &$18.89$ \\
			\hline
			\multirow{2}{*}{SVM} &Dev &$14.80$ &$15.76$\\
			&Eval &$21.34$ &$19.26$\\
			\hline
			\multirow{2}{*}{GMM} &Dev  &$16.41$ &$16.21$\\
			&Eval &$18.6$ &$18.29$\\
			\hline
			\multirow{2}{*}{DNN} &Dev  &$13.03$ &$12.92$  \\
			&Eval &$17.55$ &$15.94$ \\
			\hline
		\end{tabular}}
		\label{new_baseline_results}
\end{table}

\subsection{Model performance using endpoint detection} \label{new_benchmark_results}

We re-train all our countermeasures described in Section~\ref{our_systems} applying manual endpoint annotations during training and validation. We evaluate the performance of our new models on the development and evaluation set without using endpoint detection (test condition 1) and applying automatic endpoint detection (test condition 2). Table~\ref{new_baseline_results} summarizes the results. 

As expected our new models now show worse performance in comparison to initial models (Table~\ref{results_original_systems}) trained without endpoint detection. However it should be noted that our main focus here is not on improving EERs. We aim towards building trustworthy countermeasures providing reliable performance estimates, and making them secure from being manipulated using artefacts/cues in the dataset. Overall, deep models (CNNs and DNN) show better performance on both the development and evaluation sets compared to other countermeasures under both test conditions. This demonstrates their superiority in learning discriminative features. We find that all our models show better generalization on the evaluation set using speech endpoint detection (condition $2$). On the development set, we see a similar behaviour for GMM and DNN countermeasures but CNNs, Cosine and SVM countermeasures show slightly better performance without using endpoint detection (condition 1). Furthermore, the small performance difference of these models between test conditions $1$ and $2$ suggest that they are now less sensitive to artefacts in the dataset.

\subsection{Model robustness experiment} \label{new_model_robustness_test}

We now demonstrate the robustness of our newly trained countermeasures against being manipulated using BCS artefacts through an intervention experiment illustrated in Fig.~\ref{intervention_pipeline}. We perform this intervention on all the test recordings in the development and evaluation sets using the same $100$ ms BCS signature from Section~\ref{manipulating_models_decisions} with one major difference. Here the intervention module performs two tasks. First, it applies an automatic speech endpoint detector to remove raw samples before and after the speech utterance. Second, it appends the BCS signature at the start of the cleaned speech signal. The updated signal is then passed to the subsequent modules for feature extraction and scoring. We score them using both our newly trained models and initial models. Finally, we compare and contrast their performance in terms of EER. Table \ref{robustness_test_original_and_new_models_using_BCS} summarizes the results. Numbers shown to the left of the arrow are the results of our initial models (from Table \ref{results_original_systems}) and new models (Condition 2 of Table \ref{new_baseline_results}) before the intervention, and numbers to the right denote results after the intervention experiment.

\begin{table}[th]
\caption{Model robustness results. Numbers to the left and the right of the arrow indicate EER\% before and after the intervention on test signals using BCS signature, respectively.}

	\centering
	\scalebox{1.0}{
		\begin{tabular} {cccc}
			\hline
			 & Set & New model & Initial model \\
			\hline
			\multirow{2}{*}{CNN$_1$} &Dev  &$9.0\rightarrow10.08$ &$7.7\rightarrow34.5$  \\
			&Eval &$14.58\rightarrow18.01$ &$10.7\rightarrow36.19$ \\
			\hline
			\multirow{2}{*}{CNN$_2$} &Dev  &$9.49\rightarrow7.85$ &$7.37\rightarrow8.25$ \\
			&Eval  &$14.77\rightarrow20.96$ &$13.4\rightarrow22.6$\\
			\hline
			\multirow{2}{*}{Cosine} &Dev  &$15.91\rightarrow15.24$ &$10.6\rightarrow15.11$ \\
			&Eval &$18.89\rightarrow19.11$ &$14.8\rightarrow18.13$ \\
			\hline
			\multirow{2}{*}{SVM} &Dev &$15.76\rightarrow15.43$ &$10.8\rightarrow15.84$\\
			&Eval &$19.26\rightarrow19.33$ &$15.6\rightarrow18.84$\\
			\hline
			\multirow{2}{*}{GMM} &Dev  &$16.21\rightarrow15.50$ &$9.2\rightarrow16.85$\\
			&Eval &$18.29\rightarrow19.65$ &$13.7\rightarrow22.48$ \\
			\hline
			\multirow{2}{*}{DNN} &Dev  &$12.92\rightarrow12.40$ &$11.57\rightarrow13.33$  \\
			&Eval &$15.94\rightarrow17.91$ &$28.95\rightarrow31.46$ \\
			\hline
			
		\end{tabular}}
		\label{robustness_test_original_and_new_models_using_BCS}
\end{table}

From the increased EERs of our initial model, it is evident that the effect of this intervention on countermeasure models trained on the original training data containing BCS is much higher in comparison to the new models. We observe that our proposed frame-level DNN shows the best results under this intervention, demonstrating its robustness on this dataset. Furthermore, under the initial training conditions (without endpoint detection) the error rate of this DNN changes by about 2.5\% (on the evaluation set), and is the smallest absolute change among all other initial models including CNNs. Overall, our proposed approach of training countermeasures using endpoint detection demonstrates robust performance over the initial models. This holds true for all our models studied in this paper.  

\section{Discussion and conclusions} \label{discussion_conclusion}

Machine learning models make decisions by learning underlying patterns in the training data. As discussed in Section \ref{sec:introduction} artefacts present in a dataset can affect a wide range of ML tasks, and the impact caused by such artefacts in ASV anti-spoofing can be costly. It is therefore important to ensure that artefacts in a dataset are taken into account to build reliable ML models. To that end, this paper focus on security for voice biometric using a benchmark ASVspoof 2017 dataset which is a popular dataset for replay spoofing attack detection used in more than $60$ published research papers. Through qualitative analysis we identified artefacts on this dataset, and further investigated and confirmed their influence on models decision through intervention experiments (Section \ref{exploring_impacts_setup1}). Among different artefacts, BCS provides strong cues for the bonafide class. We also find that silence serves as a cue for the bonafide class (Section \ref{silence_cue}). Our intervention experiments using DTMF sounds show that it has no influence on model decisions. 

``Ok Google'' (S02) is the shortest duration ($0.7$ to $0.8$ seconds) phrase among the ten phrases (summarized in \cite{bhusanTASLP_supplementary}) used in the ASVspoof 2017 dataset. However, a large number of S02 examples was found to have duration more than $1.5$ seconds (see Section \ref{artefacts_in_dataset}). Therefore, we hypothesize and confirm that S02 recordings are highly affected against the attack we demonstrated in Section~\ref{manipulating_models_decisions}. We provide details in our supplementary document \cite{bhusanTASLP_supplementary}.

We emphasize two main reasons why accountability of artefacts is important while building anti-spoofing systems on this dataset. First, models trained without preprocessing these artefacts (for example BCS) may lead to overestimating the actual performance (Section \ref{exploring_impacts_setup1}). In other words, a proposed feature and a classifier might yield good replay detection performance on this dataset but might perform poorly in real-world replay detection applications. Second, they leave loopholes for being attacked by a simple copy-paste mechanism as we demonstrated in Section \ref{manipulating_models_decisions}.

To that end, we propose a method to mitigate the impact of artefacts on this dataset, and build reliable and trustworthy models. For this, a speech endpoint detection module (Fig. \ref{proposed_countermeasure_solution}) that discards every audio sample before and after the actual speech utterance is proposed. This ensures that both classes of audio now have a similar pattern, forcing learning algorithms to now focus exploiting factors of interest --- for example channel characteristics, in solving the spoofing detection problem, thus producing reliable performance estimates. We further demonstrated the effectiveness of our proposed method in mitigating the effect of dataset artefacts making countermeasures more trustworthy (Section \ref{new_model_robustness_test}). We emphasize that one of our goals in this paper is to provide awareness to the users of this dataset about potential issues and their impact on countermeasure performance. We acknowledge that ASVspoof 2017 dataset is one of the best of its kind as it contains a wide range of spoofed examples collected under real replay attack conditions. To make best use of this dataset while promoting reliability and trustworthiness, we recommend use of speech endpoint detection. Furthermore, it should also be noted that proposing novel machine learning models that are inherently robust to dataset artefacts is out of the scope of this paper. 

To conclude, replay spoofing attack detection is indeed a difficult task to solve. Artefacts in the training data can be easily exploited as we have demonstrated in this work and in \cite{bhusan2019challenge} on ASVspoof 2019 PA, the latest benchmark replay dataset. Countermeasures trained with such dataset artefacts might yield good performance as indicated by some metric (eg. EER), however their trustworthiness is called into question as they behave much like a ``horse'' in machine learning \cite{jose_clever_hans,sturm2016horse}, overestimating the actual performance. It is therefore important to ensure that both bonafide and spoof class training data have similar audio patterns and differ only in terms of channel characteristics. This encourages the development of reliable and trustworthy spoofing countermeasure models as they are now forced to exploit only the relevant factors of interest during training. 

One potential drawback of our proposed methodology is that it would not work for cases when artefacts appear within-utterance. Although in the ASVspoof 2017 dataset we did not find BCS and DTMF artefacts within utterance during our qualitative analysis, we found few utterances containing silence (and near-silence) within the utterance. Since an ASV system usually has a VAD component that discards all nonspeech samples, therefore, from a practical application perspective it would be interesting to study countermeasure design by applying a VAD on the whole utterance. We aim to extend this idea as part of our future work using the benchmark ASVspoof 2017 and ASVspoof 2019 datasets. Studies on fooling ML models using adversarial machine learning is an active research topic \cite{Nguyen_fool_dnn,douglas_fooling_AI,christian_dnn_properties}. Investigating the robustness of anti-spoofing systems against adversarial attacks would be another research avenue we look forward to. Furthermore we also aim to investigate the interpretability of such models when there are no dataset biases to uncover what attributes these models are learning to make decisions.

\ifCLASSOPTIONcaptionsoff
  \newpage
\fi

\bibliographystyle{IEEEtran}
\bibliography{mybib}

\begin{IEEEbiography}[{\includegraphics[width=1in,height=1.25in,clip,keepaspectratio]{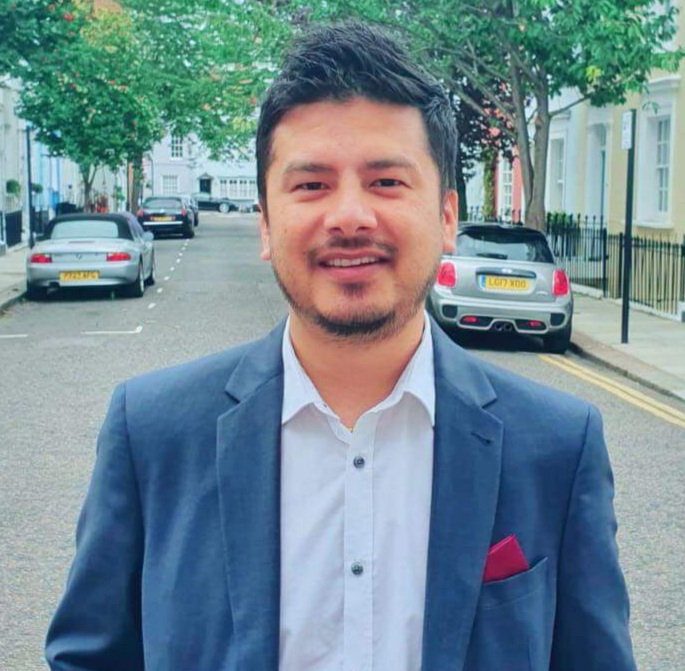}}]{Bhusan Chettri}
is a Lecturer in Data Analytics at the School of Electronic Engineering and Computer Science, Queen Mary University of London (QMUL), UK. He received the Ph.D. degree in Electronic Engineering from QMUL in 2020. During May-October 2019, he visited the Computational Speech research group, University of Eastern Finland to work on Voice Anti-Spoofing project (NOTCH) with Dr. Tomi Kinnunen. Between May-August 2020 he was a post-doctoral researcher on the NOTCH project. Having received the M.Sc degree in Speech and Language Processing from the University of  Sheffield in 2014, he worked as a Research Assistant with the Speech and Hearing Research group, Sheffield University for two years. His research focuses on machine learning and its applications to speaker recognition, spoofing and countermeasures for robust speaker recognition.
\end{IEEEbiography}

\begin{IEEEbiography}[{\includegraphics[width=1in,height=1.25in,clip,keepaspectratio]{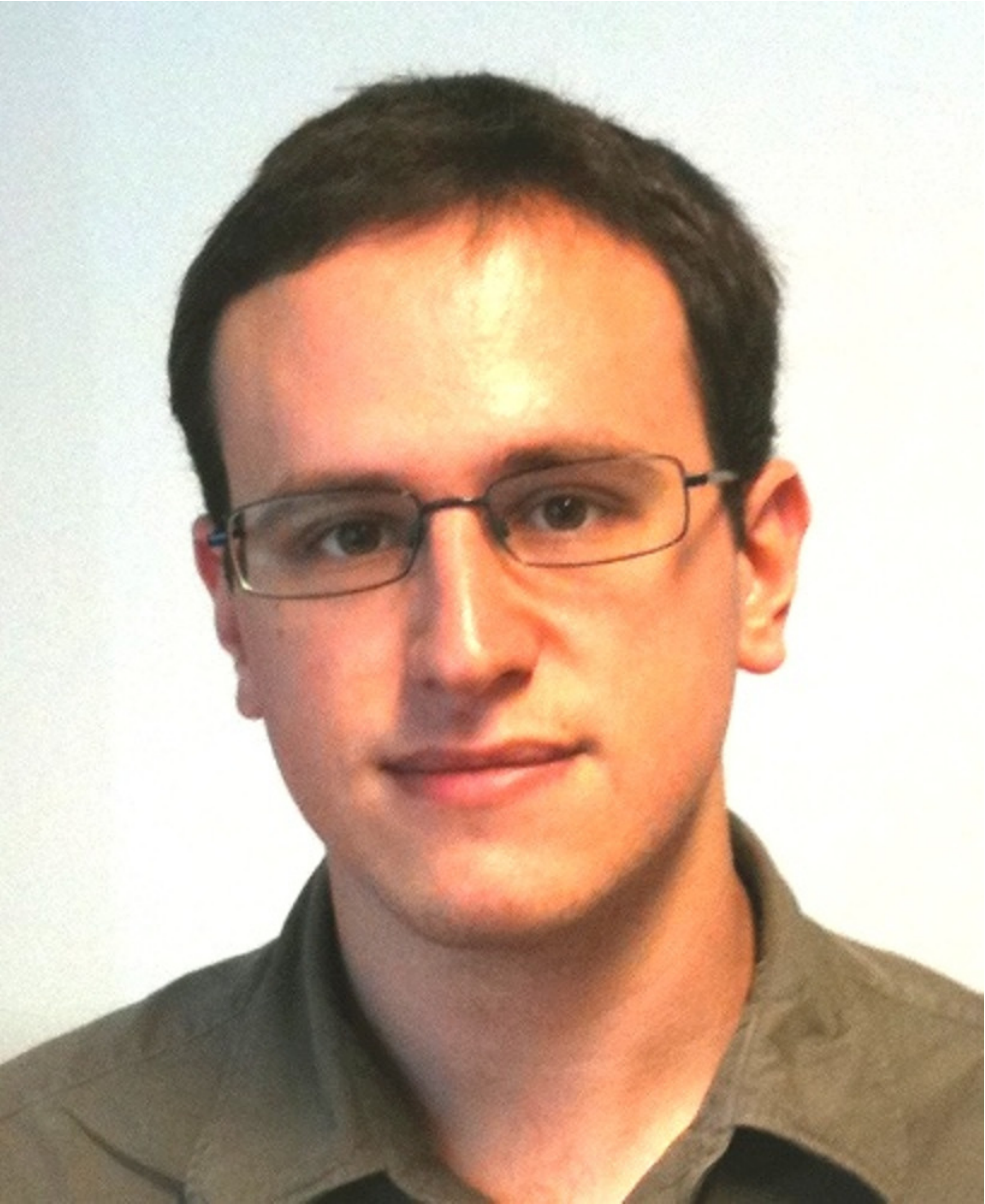}}]
{Emmanouil Benetos} (S'09, M'12, SM'20) is Senior Lecturer with the School of Electronic Engineering and Computer Science, Queen Mary University of London, and Turing Fellow with The Alan Turing Institute. He received the Ph.D.\ degree in Electronic Engineering from Queen Mary University of London, U.K., in 2012. From 2013 to 2015, he was University Research Fellow with the Department of Computer Science, City, University of London. He has authored/co-authored over 100 peer-reviewed papers spanning several topics in audio and music signal processing. His research focuses on signal processing and machine learning for audio and music signal analysis, as well as applications to music information retrieval, acoustic scene analysis, and computational musicology.
\end{IEEEbiography}

\begin{IEEEbiography}[{\includegraphics[width=1in,height=1.25in,clip,keepaspectratio]{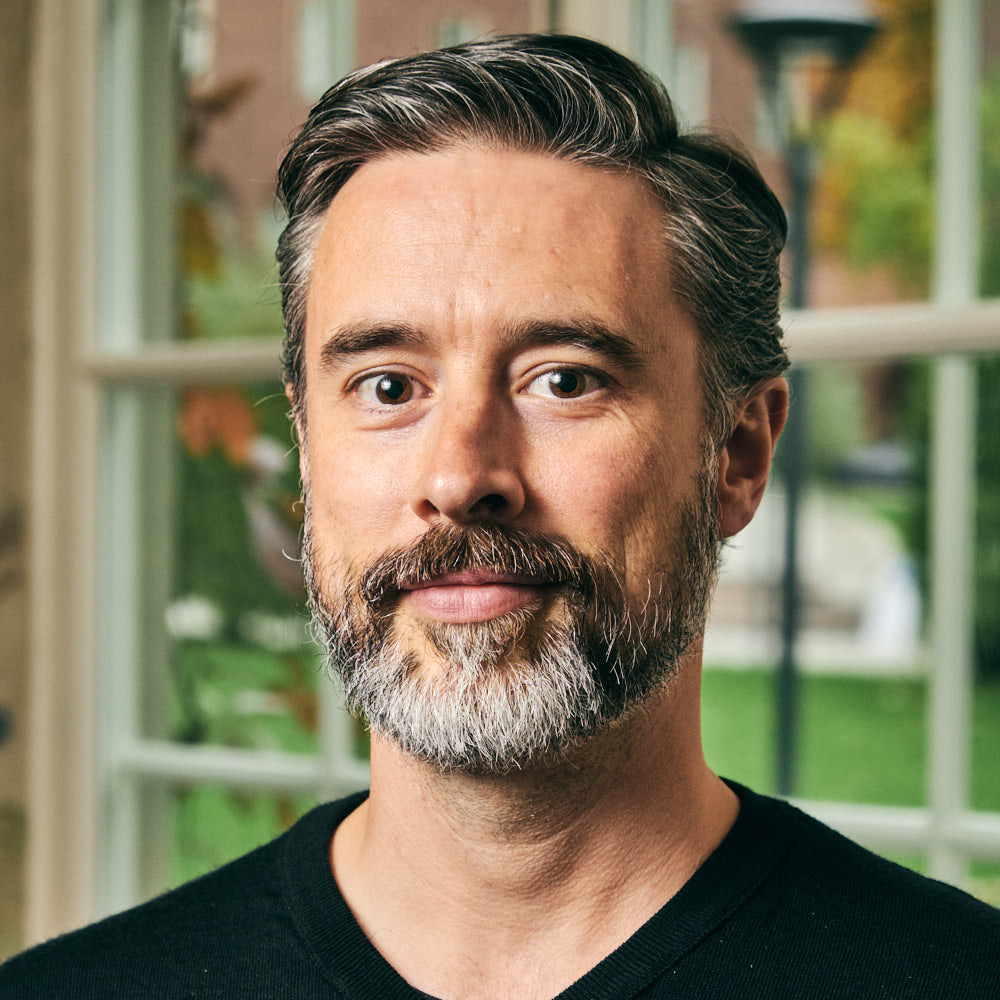}}]{Bob L. T. Sturm}
received the B.A. degree in physics from University of Colorado, Boulder in 1998, the M.A. degree in Music, Science, and Technology, at Stanford University, in 1999, the M.S. degree in multimedia engineering in the Media Arts and Technology program at University of California, Santa Barbara (UCSB), in 2004, and the M.S. and Ph.D. degrees in Electrical and Computer Engineering at UCSB, in 2007 and 2009. In Dec. 2014, he became a Lecturer at the Centre for Digital Music at Queen Mary University of London. In July 2018 he became an associate professor of computer science at the Royal Institute of Technology KTH in Stockholm Sweden. In December 2019 he was awarded an ERC Consolidator Grant for the project ``MUSAiC: Music at the Frontiers of Artificial Creativity and Criticism'' (No. 864189).
\end{IEEEbiography}
\vfill

\end{document}